\documentstyle[12pt,aasms4]{article}

\begin{document}

\title{Spectroscopic Ananlyses\\
of the Planetary System Candidates:\\
55 Cnc, 51 Peg, 47 UMa, 70 Vir, and HD 114762}
\author{Guillermo Gonzalez}
\affil{Department of Astronomy, University of Texas, Austin, TX 78712}
\authoremail{gonzalez@crown.as.utexas.edu}

\begin{abstract}
The stars 55 Cnc, 51 Peg, 47 UMa, 70 Vir, and HD 114762 have recently been 
proposed to harbor planetary mass companions.  Using spectroscopic methods we 
find that 55 Cnc and 51 Peg are super metal-rich, 47 UMa and 70 Vir have 
roughly 
solar metallicities, and HD 114762 is metal-poor.  Otherwise, the abundance 
patterns, expressed as [X/Fe], are approximately solar.  The ages of 47 UMa
 and 51 Peg are similar to that of the Sun; 70 Vir is slightly older, 
and 55 Cnc and HD 
114762 are at least 10 Gyrs old.  Our estimates of $v \sin i$ for the parent
 stars are 
1.4 $\pm$ 0.5, 1.8 $\pm$ 0.4, 2.0 $\pm$ 0.3, $<$ 1, and $<$ 1.5 km~s$^{-1}$ 
for 55 Cnc, 51 Peg, 47 UMa, 70 Vir, and HD 114762, respectively.  Using these 
data and estimates for the rotation periods and radii, the corresponding
 masses of 
the companions are: $>$ 0.6, $0.51^{+0.22}_{-0.03}$, 4.6 $\pm$ 1.0, $>$ 20, 
and $>$ 8.5 M$_{\rm J}$.  The systems appear to fall into three groups: 
roughly 
Jupiter mass companions with small circular orbits with metal-rich parent 
stars (55 
Cnc and 51 Peg), larger companions with larger circular orbits (47 UMa), and 
much more massive companions with large eccentric orbits orbiting moderately 
metal-poor parent stars (70 Vir and HD 114762).

Our findings are consistent with a recently proposed mechanism whereby a gas 
giant migrates to within a few hundredths of an AU of its parent star during
 the 
formative epoch of the planetary system.  During this process the material
 between 
the giant planet and the star is accreted onto the latter.  If the accreted
 material is 
depleted in H and He, then the photospheric composition of the parent star 
might be 
altered significantly.
\end{abstract}


\section{Introduction}

Between October 1995 and April 1996 the discovery of four candidate 
extra-solar 
systems (55 Cnc, 51 Peg, 47 UMa, and 70 Vir) were announced (Mayor \& Queloz 
1995; Butler \& Marcy 1996; Marcy \& Butler 1996a); another candidate, HD 
114762, was discovered by Latham et al. (1989) .  These historic discoveries
 are 
the fruits of about 15 years of searching by several groups (Campbell et al.
 1988; 
Cochran \& Hatzes 1995; Walker et al. 1995), most of which have not yet found 
independent evidence of planetary mass companions.  Now, theoreticians have 
empirical data to test their planet formation models.  One aspect of planet
 formation 
that has not been discussed in much detail is the dependence of planet
 formation on 
the metallicity of the protostellar cloud.  With five planetary system
 candidates (and 
one certain case, the Solar System), we can begin to address this question.

The primary purpose of this study is to accurately determine the metallicities
 of the 
photospheres of 55 Cnc, 51 Peg, 47 UMa, 70 Vir, and HD 114762.  We also 
attempt to estimate the abundances of about 15 other elements including
 lithium, 
which can to used to constrain their ages.  While others have performed
 abundance 
analyses on these stars already, we analyze them together as a homogenous
 group; 
our analysis also serves as a check of previous studies.  As a secondary goal,
 we 
also derive accurate estimates of $v \sin i$ for these stars, and, combining
 them 
with other data, use them to place constraints on the masses of the companions.

\section{Observations}

Moderate resolution spectra ($R \approx 60000$) were obtained by the author on 
five observing runs of two to three nights each between December 1995 and May 
1996 at the McDonald observatory using the 2.1 m telescope equipped with a 
Cassegrain echelle and a 1200 x 400 pixel Reticon detector (McCarthy et al.
 1993).  
Each star was observed no more than once per night with exposure times
 typically 
between one and five minutes, resulting in spectra with S/N ratios between 150
 and 
300.  These spectra will be used in \S 3.1 for the purpose of deriving
 abundance estimates for the program stars.

For the purpose of studying line profile shapes, additional spectra were
 obtained at 
the author's request by David L. Lambert and Eric Bakker using the 2.7 m 
telescope equipped with the 2dcoud\'{e} echelle giving spectral resolutions 
(depending on the slit width) near 230000 for 47 UMa, 70 Vir, and HD 114762, 
115000 for 51 Peg, and 180000 for 55 Cnc (instrument described by Tull et al. 
1995); the S/N ratios are in the range of 250 to 350, except for HD 114762,
 which 
is near 200.  The (incomplete) spectral coverage is 4700 - 5800 \AA\ for 47
 UMa, 
70 Vir and HD 114762 and 5700 - 7900 \AA\ for 51 Peg.  Each observation was 
broken up into two or three exposures in order to detect and eliminate cosmic
 ray 
hits on the images.  The procedures followed in reducing the spectra are
 described 
in Gonzalez \& Lambert (1996).  A sample spectrum of 51 Peg overplotted
 with the solar spectrum is shown in Figure 1.

\section{Analysis}
\subsection{Abundance Analysis}

Our method of abundance analysis closely follows that used by Gonzalez \& 
Lambert in their study of several stars in the $\alpha$ Per cluster.
  Consequently, 
this analysis is a differential one with respect to the Sun.  Since the stars
 on our 
program are similar to the Sun in their physical characteristics, a
 differential 
abundance analysis with the Sun as the source of the $gf$-values avoids many 
possible systematic errors.  Two of these include uncertainties in the
 treatment of 
the structure of the atmospheres, such as differences in the mixing length to
 scale height ratio or convection and differences in the area covered by 
sunspots \footnote{The sunposts may vary on a monthly or yearly basis.  Also,
 the area covered by sunspots averaged over a sunspot cycle may vary from star
 to star; 
this may not be a problem for solar type stars younger than the Sun typically 
display stronger chromospheric activity.}.  Following Gonzalez \& Lambert we 
employ the Kurucz (1992) model atmospheres; the effective temperature, $T_{\rm 
eff}$, surface gravity, $g$, and depth-independent microturbulence parameter, 
$\xi_{\rm t}$, are estimated in the standard way using the Fe I and Fe II lines 
(details are discussed in the next section).

The abundance analyses of 55 Cnc, 51 Peg, 47 UMa, 70 Vir, and HD 114762 are 
based on measurements of the moderate resolution spectra obtained with the
 2.1 m telescope.  The linelist used in the present analysis was selected from
 the lines 
published in Tables 12 and 13 by Gonzalez \& Lambert.  Additional lines were 
added (Table 1) and their $gf$-values estimated using the equivalent widths
 (EW's) 
measured on the Solar Flux Atlas (Kurucz et al. 1984).  Although the wavelength 
coverage is quite extensive and the quality of the spectra is high, we have
 been very restrictive in our selection of lines.  To be suitable, a line must
 be unblended (except 
for a few rare cases when the line is critical to estimating the abundance
 of a given 
element) and have a symmetric profile in both the target star spectrum and the
 solar 
spectrum.  The sample is restricted to mostly moderate strength lines between
 about 
20 and 150 m\AA.  Weaker lines have a larger relative error in the EW estimates 
due to noise, and stronger lines are more sensitive to errors in $\xi_{\rm t}$; 
based on measurements of the same lines on spectra taken on different nights,
 we 
estimate that the average uncertainty in an EW estimate to be about $\pm$ 1-2 
m\AA\ for lines with EW = 20 - 60 m\AA\ and about $\pm$ 2-3 m\AA\ for stronger 
lines.  Such low errors were achieved by smoothing isolated lines (effectively 
increasing the S/N ratio), measuring some lines on the overlap regions of
 adjacent 
orders, and averaging EW measurements obtained on different observing runs.  
The range in measured line strengths is sufficient to estimate $\xi_{\rm t}$ 
accurately.  The Fe I lines employed in the analysis have lower excitation 
potentials,$\chi_{\rm l}$, ranging from 2.2 to 5.0 eV; this is sufficient to
 estimate $T_{\rm eff}$ accurately for each of the program stars.

\subsubsection{Model Atmosphere Selection and Fe Abundances}

While hundreds of Fe lines are present in the spectral regions observed, we
 selected only 26 Fe I and 5 Fe II for use in the analysis (the total number
 varies from star to 
star).  The values of $T_{\rm eff}$, $\log g$, and $\xi_{\rm t}$ have been 
estimated with this sample of Fe lines (Table 2).  The typical uncertainties
 in these parameters are $\pm$ 100 K, $\pm$ 0.1 (cgs), and $\pm$ 0.1
 km~s$^{-1}$, respectively; the individual uncertainties are listed in Table 2.
  They lead to a typical 
uncertainty in [Fe/H] of $\pm$ 0.07 dex, which was calculated using the 
sensitivities of the abundances to changes in atmospheric parameters (Table 3).
  The sensitivities of the $\log \epsilon_{\rm Fe}$ versus $\chi_{\rm l}$ and
 $\log \epsilon_{\rm Fe}$ versus $\log \left(EW/\lambda\right)$ relationships
 to changes in $T_{\rm eff}$ and $\xi_{\rm t}$, respectively, are shown in
 Figure 2 for 51 Peg.  The measured EW's are listed in Table 4.

How do our estimates of [Fe/H] and $T_{\rm eff}$ compare to published 
estimates?  They can be calculated from Str\"omgren $uvby-\beta$ narrow band 
photometry; such data exist for all the program stars (Hauck \& Mermilliod
 1990) 
\footnote{The quantity [Me/H] refers to the logarithmic abundance of a metal 
relative to the sun.  Technically, one actually measures [Me/H] with photometry 
rather than [Fe/H], since the spectral regions covered by the different
 Str\"omgren 
filters includes other elements in addition to iron.}.  Using the calibration
 equation between [Me/H] and the $\beta$ and dm1 indices given by Nissen
 (1988), we estimate that [Me/H] is 0.27, 0.03, -0.06, and -0.79 for 51 Peg,
 47 UMa, 70 Vir, and HD 114762, respectively (these photometric calibrations
 do not apply to 55 Cnc as it is too cool); Nissen claims his equation is
 accurate to within about $\pm$ 0.10 dex.  Thus, given the uncertainties of
 our spectroscopic estimates, these photometric estimates are consistent with
 ours.  Saxner \& Hammarb\"ack (1985) derived equations relating the $b-y$ 
and $\beta$ indices to $T_{\rm eff}$.  While their calibrations are accurate
 to with about $\pm$ 60 K, we can improve upon them 
slightly by updating the photometry (using the mean values tabulated by
 Hauck \& 
Mermilliod 1990) of the calibrating stars and restricting the sample to sharp-
lined stars.  With a sample of 16 stars, we derive,

\begin{eqnarray}
T_{\rm eff}=8178-5980(b-y)\{1-0.070[Fe/H]\} \\
T_{\rm eff}=11495\sqrt{(\beta-2.3405)}\nonumber
\end{eqnarray}

The mean scatter in the first and second equations are $\pm$ 40 and $\pm$ 50 K, 
respectively.  We list in Table 5 the resultant estimates of $T_{\rm eff}$ for
 each of 
our program stars using equation 1 and the mean Str\"omgren indices from Hauck 
\& Mermilliod.  Also listed in the table are the mean values of $T_{\rm eff}$
 for each program star combining the photometric and spectroscopic
 estimates and the resultant corrected values of log $g$ and [Fe/H] for each
 star.  These new atmospheric parameters will be used in \S 4.2 to estimate
 the ages of the program stars.

\subsubsection{Other Elements}

The abundances of 15 additional elements were estimated using the atmospheric 
parameters given in Table 2.  The individual line measurements are listed in
 Table 4.  The uncertainties in the [X/H] values were calculated using the
 estimated uncertainties in the atmospheric parameters (Table 2) and the
 data in Table 3.

Since the resonance line of lithium is blended with other lines in spectra
 of solar-type stars, we must employ spectrum synthesis methods to estimate
 its abundance.  
We employed the linelist given by Cunha et al. (1995, Table 7), modified
 slightly to 
reproduce the solar spectrum using the Kurucz solar model atmosphere (Kurucz 
1992); the spectral region synthesized spans 6700 to 6710 \AA\.  The same
 stellar 
atmospheric parameters derived from the Fe-line analysis were used in producing 
the synthetic spectra.  The line broadening was approximated with a Gaussian 
function in the synthetic spectra with a width chosen so that the two strong
 Fe I lines in the observed spectra are reproduced accurately.  The lithium
 line is discernible by eye on all spectra, except that of 55 Cnc, where it
 is not detectable at the level of the noise.  The final adopted lithium
 abundances for the three stars are listed in Table 6 along with the other
 elements.

\subsection{$v \sin i$}

In order to estimate the mass of a companion, we require, among other
 quantities, 
an estimate of the orbital inclination.  Assuming the orbital axis is aligned
 with the 
stellar rotation axis, an estimate of the projected stellar equatorial
 rotational velocity 
($v \sin i$) can be used to constrain the orbital inclination.  In this
 section we 
describe the derivation of $v \sin i$ for the program stars from the high
 resolution 
spectra obtained with the 2.7 m telescope.  In addition to the program star
 spectra, a 
spectrum of the sky was obtained in order to calibrate our technique with the
 known solar parameters.

Of the two methods most often used to estimate $v \sin i$ from a stellar
 spectrum, 
Fourier transform and profile synthesis, we opted for the latter.  The
 instrumental, 
macroturbulent ($\zeta_{\rm RT}$), microturbulent ($\xi_{\rm t}$), rotational,
 and 
thermal broadening mechanisms are included in the analysis of the line profiles.  
The syntheses have been carried out with MOOG, modified to include 
macroturbulent line broadening in addition to the other line broadening
 mechanisms 
already present in the program.  We have adopted the radial-tangential
 description 
of macroturbulent line broadening (Gray 1992) using the mathematical formalism
 of 
Durrant (1979).  The instrumental broadening is approximated by a Gaussian 
function, its width determined from the Th-Ar comparison spectrum obtained 
immediately following each stellar observation.  This approximation is a very
 close 
fit to the Th-Ar lines; however, even if it were not, that would not cause
 significant 
errors since the instrumental broadening is relatively small compared to the
 other line broadeners.  The limb darkening coefficients, required for
 synthesizing the rotational profiles, are interpolated from Figure 17.6 of
 Gray (1992).  The model atmospheric parameters used in the syntheses are the
 same as those estimated from the Fe-line analyses in \S 3.1.1, except for
 $\xi_{\rm t}$.  In their studies of the solar spectrum Gray (1977) and Takeda
 (1995) found that one must assume $\xi_{\rm t}$ = 0.5 km~s$^{-1}$ in order to
 accurately reproduce the line profiles accurately.

The Fe I lines at 5379.586 and 5638.249 \AA\ were selected for analysis from 
Table 2 of Takeda; these lines are unblended, are moderate in strength, and
 have smaller than average $\zeta_{\rm RT}$ values.  This last criterion is
 particularly important in deriving accurate values of $v \sin i$, since
 $\zeta_{\rm RT}$ is the 
dominant broadening mechanism in our sample stars.  The solar Fe I lines were 
analyzed first with $v \sin i$ held fixed at 1.9 km~s$^{-1}$ in order to
 estimate 
$\xi_{\rm t}$ and $\zeta_{\rm RT}$.  The parameters were adjusted manually and 
the final parameters chosen when the residuals between the observed and
 synthetic 
line profiles were minimized.  The value of $\xi_{\rm t}$ required to reproduce
 the 
line profiles is 0.4 km~s$^{-1}$; our estimates of $\zeta_{\rm RT}$ are similar
 to 
those quoted by Takeda for the same lines.  Treating $v \sin i$ as an
 additional free parameter leads to a very similar solution (Table 7).  The
 apparent discrepancy in $\xi_{\rm t}$ obtained from the abundance analysis as
 compared to the line profile 
analysis has been noted by other researchers.  However, it is only a scientific 
problem, not a practical one.  The discrepancy is likely caused by model-
incompleteness (cf. Takeda et al. 1996 for a brief discussion of this problem).

The values of $v \sin i$ for the other program stars were estimated for the
 program stars in the same way as for the Sun.  Zeeman broadening was not
 included in any of the analyses, since it is not found to be a significant
 source of line broadening for stars hotter that G6 (Gray 1984).  This may not
 be the case for 55 Cnc, but as we show in \S 4.2, its low chromospheric
 activity implies that Zeeman broadening may 
not be significant for this star.  The values of $v \sin i$ and
 $\zeta_{\rm RT}$ for the program stars were estimated assuming
 $\xi_{\rm t}$ = 0.4 km~s$^{-1}$, except for 47 UMa, where $\xi_{\rm t}$ =
 0.45 km~s$^{-1}$ was assumed.  The estimated values for each spectral line
 are given in Table 7.  The uncertainties 
quoted were estimated in a formal way by noting the change in residuals as
 the broadening parameters are changed.  Taking into consideration Soderblom's
 (1982) conclusion that small $v \sin i$ estimates, those less than about 2
 km~s$^{-1}$, should be regarded as upper limits, we adopt the following
 values of $v \sin i$: 1.4 
$\pm$ 0.5, 1.8 $\pm$ 0.4, 2.0 $\pm$ 0.3, $<$ 1, and $<$ 1.5 km~s$^{-1}$ for 
55 Cnc, 51 Peg, 47 UMa, 70 Vir, and HD 114762, respectively.  Slight 
asymmetries were detectable in some of the lines, but they did not affect the
 final solutions.  A sample line profile fit is shown in Figure 3.

Previous estimates of $v \sin i$ exist for 51 Peg and HD 114762.  The values
 of $v \sin i$ or 51 Peg are 2.1 $\pm$ 0.6 (Baranne et al. 1979), 1.7 $\pm$
 0.8 (Soderblom 1983), 2.8 $\pm$ 0.5 (Mayor \& Queloz 1995), and 2.4 $\pm$ 0.3 
km~s$^{-1}$ (Fran\c{c}ois et al. 1996).  Cochran et al. (1991) derived
 $v \sin i$ = $0^{+1}_{-0}$ km~s$^{-1}$ for HD 114762 assuming $\xi_{\rm t}$ =
 1.0 km~s$^{-1}$, while Hale (1995) found $v \sin i$ = 0.8 $\pm$ 0.7
 km~s$^{-1}$ assuming $\xi_{\rm t}$ = 0.7 km~s$^{-1}$.

\section{Discussion}
\subsection{Abundances}

All our program stars are listed in the catalog of Cayrel de Strobel et al.
 (1992), who compile spectroscopic [Fe/H] determinations from the literature.
  They list five estimates for HD 114762 averaging to -0.77, which is 0.16 dex
 less than our estimate.  For the other stars all the estimates are from the
 1970's: there are four for 55 Cnc averaging to +0.13; 51 Peg, 47 UMa, and 70
 Vir each have single estimates of +0.12, -0.02, and -0.11, respectively.
  Mayor \& Queloz (1995) quote a recent (previously unpublished) spectroscopic
 estimate of [Fe/H] for 51 Peg of 0.19 dex 
and a Geneva photometric estimate of [Me/H] of 0.20 dex.  Edvardsson et al. 
(1993) included 51 Peg, 47 UMa, and HD 114762 in their large survey of 189 
nearby F and G dwarfs obtaining [Fe/H] values of +0.06, +0.01, and -0.74, 
respectively; these estimates are systematically smaller than ours by
 about 0.13 dex.

The photometric metallicity estimates of Edvardsson et al. are systematically
 higher than their spectroscopic estimates for the most metal-rich stars in
 their sample.  
Nissen (1994) addressed this discrepancy and claims that the problem is with
 the photometric metallicity calibration used by Edvardsson et al.  In
 order to check the possibility that the different [Fe/H] estimates for 51 Peg
 are due to differences in the 
EW measurements, we have compared the Edvardsson et al. measurements (not 
included in their published paper; provided to the author by Bengt Edvardsson)
 to those of this study.  The EW measurements of Edvardsson et al. are
 systematically smaller by about 15\%; this is sufficient to account for the
 different [Fe/H] estimates 
of our studies.  Our measurements, however, are very similar to those obtained 
recently by Jocelyn Tomkin, who has recently observed 51 Peg with the 2.7 m 
telescope at McDonald Observatory with a spectral resolution $\approx 60000$.  
The spectral resolution used in the Edvardsson et al. (1993) study was only 
$\approx 30000$.  Such a low resolution, especially for a metal-rich star,
 might 
result in blending of the numerous weak lines, leading to the formation of a
 pseudo-continuum and apparent weakening of the stronger lines.

In his study of super metal-rich (SMR; defined as having [Fe/H] $>$ 0.2 ) stars 
Taylor (1996; his Table 4) lists 29 SMR luminosity class IV-V candidates;
 55 Cnc and 51 Peg are included in this list.  Photometric estimates of [Fe/H]
 are +0.5 and 
+0.25 for 55 Cnc and 51 Peg, respectively; mean spectroscopic [Fe/H] values, 
based on published estimates and transformed to a uniform temperature and 
metallicity scale by Taylor, are 0.41 $\pm$ 0.10 and 0.17 $\pm$ 0.05, 
respectively.  While having two of our program stars appear in a list of 29 
suspected SMR stars seems significant, it becomes less so when one realizes
 that 10 of the nearly 120 stars analyzed by Marcy and Butler also appear on
 the list.  
More significant is the fact that 55 Cnc is one of the 7 stars with a probability
 $ge$ 
95\% of being a SMR star; none of the other stars observed by Marcy \& Butler
 is a 
member of this "magnificent seven."  Including the results of our study and those 
of Mayor \& Queloz (1995) leads us to conclude that 51 Peg is marginally also a 
SMR star.

As an illustration of the unusual metallicity distribution of our program stars,
 we 
compare them to the distribution of [Fe/H] values for nearby F and G dwarfs.  
Marsakov \& Shevelev (1995) have estimated the [Fe/H] distribution from $uvby$ 
photometry for 5489 F to early G dwarfs within 80 parsecs of the Sun.  Their 
method involves using the $b-y$ color index as the independent variable in 
estimating [Fe/H], which is slightly less precise than the traditional method
 using 
the Str\"omgren $\beta$ index as the independent variable.  We have edited
 their 
dataset by removing unresolved binaries and stars with $T_{\rm eff}$ $>$
 6500 K, 
leaving 3552 stars in the sample (Figure 4).  The mean value of [Fe/H] for this 
distribution is -0.16 dex.  This result is similar to that of the more careful
 analysis 
of a smaller set of nearby dwarfs by Rocha-Pinto \& Maciel (1996).  As we will 
show below, the companions to 70 Vir and HD 114762 are not likely to be 
planetary in nature.  This leaves 55 Cnc, 51 Peg, and 47 UMa as the best 
candidates harboring planets, which also happen to be the most metal-rich stars
 in our sample.

We do not confirm the finding of Edvardsson et al. (1993) that 51 Peg is a 
"NaMgAl" star.  Apart from the general enhancements of the metals, the
 abundance 
patterns of the program stars are very similar to that of the Sun.  To the best
 of our 
knowledge, the Li abundance has only been estimated for one of our sample
 stars, 51 Peg;  Fran\c{c}ois et al. (1996) find that it has a solar Li
 abundance.  The relevance of knowledge of the lithium abundances will be
 discussed in the next section.

\subsection{Ages, Rotation Periods, and Masses}

The $v \sin i$ estimates, when combined with the rotational velocities and mass 
estimates of the program stars, can be used to set limits on the masses of
 their 
companions.  Given that angular velocity correlates fairly well with age for G 
dwarfs (Dorren et al. 1994; Soderblom 1985), age can be used to place useful 
constraints on the rotation periods for our sample.  In this section we will
 estimate 
the ages, rotation velocities, and masses of the program stars as well as the
 masses 
of the companions using the atmospheric parameters derived in \S3 and published 
data.

The most commonly used method of estimating the mass and age of a single main 
sequence or subgiant star involves comparing its observed physical parameters 
($T_{\rm eff}$, $M_{\rm V}$, [Fe/H]) to theoretical stellar evolutionary 
sequences.  We make use of the VandenBerg (1985) evolutionary stellar grids, 
extrapolated slightly to the metallicities of the program stars (VandenBerg did not 
calculate tracks above solar metallicity), to estimate their ages and masses.  We
 first 
checked the accuracy of VandenBerg's calculations with the Sun.  The theoretical 
tracks predict the correct value of $T_{\rm eff}$ for the Sun given its mass, age, 
and metallicity, but the theoretical value of and $M_{\rm V}$ must be reduced
 by 
0.21 dex.  We have derived two estimates of $M_{\rm V}$ for each star: one uses 
the HIPPARCOS parallax estimates of Perryman et al. (1996), except for 55 Cnc, 
which is from van Altena et al. (1995) corrected according to Lutz \& Kelker 
(1973), the other uses the Strömgren photometric indices and equations
 10 and 12 
of Nissen (1988).  The $M_{\rm V}$ estimates from the HIPPARCOS data are 
much more accurate; they were adopted instead of the photometric estimates when 
available.  We list the ages in Table 8, which were calculated using the atmospheric 
parameters given in Table 5 and the adopted $M_{\rm V}$ estimates, corrected 
using the factor derived from the solar case above.  We can check the accuracy of 
our method with a star of known age; $\beta$ Hyi is probably the best choice as it
 is the closest subgiant and has nearly the same temperature as the Sun.  Using the 
physical stellar parameters of $\beta$ Hyi given in Dravins et al. (1993), we derive 
an age of 12.5 $\pm$ 2.5 Gyr and a mass of 0.94 $\pm$ 0.04 M$_{\odot}$. 
Dravins et al. derived an age of 9.5 $\pm$ 0.8 Gyr and a mass of 0.99 
M$_{\odot}$, consistent with our estimates.  Edvardsson et al. (1993) derived age 
estimates for 51 Peg, 47 UMa, and HD 114762 also using the VandenBerg tracks; 
given the uncertainties, their results are generally in agreement with ours
 except for 
51 Peg.  This is not unexpected given our use of a more accurate estimate for 
$M_{\rm V}$ for this star.

There are other, less precise, age indicators that can be used to further
 constrain the 
ages of the stars.  Several indicators have to do with the Galactic orbit.
  As a star 
ages, chance encounters with molecular clouds perturb it away from a simple 
circular orbit in the plane of the disk.  Hence, both the maximum height
 reached 
above the Galactic plane, $Z_{\rm max}$, and the eccentricity,$e$, of its orbit 
generally increase with time.  Among our sample, HD 114762 has the largest 
values of $e$ and $Z_{\rm max}$, 0.27 and 0.90 kpc, respectively.  Given the 
large scatter in the $Z_{\rm max}$ - age and $e$ - age relations (see
 Edvardsson et 
al. 1993), we cannot say much more about the other stars, except that 47 UMa is 
likely the youngest.  The total space velocity is another age indicator; the 
magnitudes of the space velocities for 55 Cnc, 51 Peg, 47 UMa, 70 Vir, and HD 
114762 are 42.1, 38.0, 24.2, 26.9, and 104.8 km~s$^{-1}$, respectively.  Using 
the young-disk versus old-disk classification scheme of Eggen (1969; his Figure 1) 
based on $U$ and $V$ velocities, only HD 114762 clearly stands out as a member 
of the old-disk.  Better age indicators are those related to the level of
 chromospheric 
activity.  As a star ages, its chromospheric activity is seen to decline
 (Soderblom 
1985; Dorren et al. 1994).  Of our sample, all but HD 114762 were included in 
Soderblom's study of Ca II emission strengths in solar-type dwarfs; of the 
remaining four stars, 47 UMa is found to be the youngest and 55 Cnc the oldest.  
Related to the level of chromospheric activity is the amplitude of the visual 
brightness; by the time a solar-type dwarf is about 3 Gyr old, the brightness 
variations drop below about 0.01 magnitudes (Dorren et al. 1994).  Shortly
 after 
the discovery of 51 Peg B was announced in October 1995, 51 Peg was monitored 
photometrically and was found to be constant at the 0.0018 mag. level (Guinan 
1995), implying that 51 Peg is at least a few billion years old.  Finally, the
 lithium 
abundance might be used to constrain stellar age.  This is more tricky, though, 
since the decline of lithium in a star's photosphere as it ages is followed
 eventually 
by a dramatic resurgence as the star ascends the subgiant branch (Dravins
 et al. 
1993); it is somewhat variable from cluster to cluster.  While there has been
 some 
progress in deriving the relation between age and lithium abundance for dwarfs 
(Soderblom et al. 1990), more recent work has shown that the relation is not be 
simple one (Lambert et al. 1991; Favata et al. 1996).  Hence, while a large
 lithium 
abundance cannot tell us much by itself, a very low abundance is consistent
 with an 
intermediate age (neither very young on the ZAMS nor very old on its way up the 
subgiant branch).  Using this criterion combined with our surface gravity
 estimates, 
it appears that HD 114762 is very old followed by 70 Vir and 55 Cnc.  Combining 
all these secondary age criteria, the ages of the program stars, in descending
 order, 
are HD 114762, 70 Vir, 55 Cnc, 51 Peg, and 47 UMa.  This order is similar, but 
not identical, to the order derived from the stellar evolutionary tracks.

The masses derived from the VandenBerg tracks are given in Table 10; the
 stellar 
radii, based on our estimates of $M_{\rm V}$ and $T_{\rm eff}$, are 1.03 $\pm$ 
0.08, 1.14 $\pm$ 0.06, 1.20 $\pm$ 0.06, 1.83 $\pm$ 0.10, and 1.06 $\pm$ 0.21 
R$_{\odot}$ for 55 Cnc, 51 Peg, 47 UMa, 70 Vir, and HD 114762, respectively.  
The uncertainties associated with the mass estimates may seem small, but for stars 
at or near the turnoff (such as 55 Cnc, 70 Vir, and HD 114762), the age and
 mass 
estimates are well-constrained by the observations; also, the mass and age estimates 
are well constrained with the new HIPPARCOS data.  Using Soderblom's (1985) 
estimates of angular rotation velocities, which he derived from $R_{\rm HK}$, the 
ratio of the mean Ca II H and K flux to the stellar bolometric flux, and an $R_{\rm 
HK}$ - angular velocity relation, we derive rotation periods of 44 $\pm$ 2, 29 
$\pm$ 1, 16 $\pm$ 1, and 36 $\pm$ 2 days for 55 Cnc, 51 Peg, 47 UMa, and 70 
Vir, respectively.  Using Soderblom's equation 2 with his angular velocity 
estimates, the ages of the stars are 5.6 $\pm$ 0.6, 7.6 $\pm$ 1.7, 1.4 $\pm$ 0.4, 
and 18 $\pm$ 6 Gyr, respectively.  These estimates are not in agreement with
 the 
ages derived earlier from the theoretical evolutionary tracks, which is not
 surprising 
given the indirect nature of the angular velocity - age relation.  The rotation
 period 
has not been measured for HD 114762, but we will adopt a value of 45 $\pm$ 5 
days based on its age.  Baliunas et al. (1996), using Ca II H and K flux modulation 
observations, directly measured the rotation period of 51 Peg to be 37 days (they 
did not quote error bars).  This apparent discrepancy between the Baliunas et al. 
and Soderblom rotation period estimates will need to be resolved in the future, but 
for now we will adopt 35 $\pm$ 2 days as the rotation period of 51 Peg.

Using the adopted values of $v \sin i$, rotation period, and radius for each star,
 we 
estimate the following values of $\sin i$:  1.18 $\pm$ 0.44, 1.09 $\pm$ 0.26, 0.53 
$\pm$ 0.09, $<$ 0.39, and $<$ 1.26, for 55 Cnc, 51 Peg, 47 UMa, 70 Vir, and 
HD 114762, respectively.  Obviously, values of $\sin i$ $>$ 1.0 are not acceptable; 
taking this into account, the new estimates become: $1.00^{+0.00}_{-0.28}$, 
$1.00^{+0.00}_{-0.17}$, 0.53 $\pm$ 0.09, $<$ 0.39, and $<$ 1.00.  One 
additional source of uncertainty needs to be included in the $\sin i$ estimates - the 
degree of alignment between the orbital axis of the companion and rotational axis of 
the parent star, which we will represent as $i_{\rm orb}-i_{\rm rot}$ = $\delta i$.  
Hale (1994) estimated that $\delta i$ for solar-type binaries with semimajor axes 
less that 15 AU is about 10 degrees.  Perhaps more relevant to the present analysis 
is the fact that the value of $\delta i$ for Jupiter is only about 6 degrees.
  Including 
the additional uncertainty in the orbital inclination of $\pm$ 8 degrees, the
 values of 
$\sin i$ of the companions' orbits become $1.00^{+0.00}_{-0.38}$, 
$1.00^{+0.00}_{-0.26}$, $0.53^{+0.11}_{-0.12}$, $<$ 0.39, and $<$ 1.00.

Mayor \& Queloz (1995) quote a mass function of (0.91 $\pm$ 0.15) x $10^{\rm 
-10}$ M$_{\odot}$ for 51 Peg; Butler \& Marcy (1996) and Marcy \& Butler 
(1996a) quote (1.07 $\pm$ 0.23) x $10^{\rm -8}$ and (2.98 $\pm$ 0.20) x 
$10^{\rm -7}$ M$_{\odot}$ for 47 UMa and 70 Vir, respectively.  Marcy \& Butler 
(1996b) quote M$\sin i$ = 7.6 x $10^{\rm -4}$ M$_{\odot}$ for 55 Cnc B (no 
error bars given).  Cochran et al. (1991) determine a mass function of (1.6 $\pm$ 
0.1) x $10^{\rm -7}$ M$_{\odot}$ for HD 114762.  Using these data and the 
values of $v \sin i$ quoted above we estimate the masses of the companions
 to be 
$>$ 0.6, $0.51^{+0.22}_{-0.03}$, 4.6 $\pm$ 1.0, $>$ 20, and $>$ 8.5 
M$_{\rm J}$ for 55 Cnc, 51 Peg, 47 UMa, 70 Vir, and HD 114762, respectively.

Now that we have estimates of the masses of the program stars' companions,
 their 
true nature can be constrained.  The systems can be divided into three groups: 55 
Cnc and 51 Peg with large metallicities, small eccentricities, small orbital radii,
 and 
companion masses near 1 M$_{\rm J}$, 47 UMa with a smaller metallicity, a
 larger 
circular orbit, and a larger companion mass, and 70 Vir and HD 114762 with low 
metallicities, large eccentricities, moderate orbital radii, and much larger
 companion 
masses.  It is likely that 70 Vir B and HD 114762 B are brown dwarfs or maybe 
even very low mass M dwarfs, but the uncertainties do not rule out the possibility that they are massive Jovian-type planets.

Of relevance to 51 Peg is the fact that tidally locked close binaries (orbital 
periods of a few days) have lithium abundances larger than their single star 
counterparts (Soderblom et al. 1990; Balachandran et al. 1993).  If, contrary
 to our findings, the 51 Peg system has a very small orbital inclination and
 its companion is of stellar mass, then its lithium abundance would likely
 have been larger than we measured.  Also, the rotation period of 51 Peg does
 not allow the possibility that this is a tidally locked system (Mayor \&
 Queloz pointed out that the lack of synchronization on Gyr timescales is a
 strong argument against a stellar mass companion for 51 Peg).

\subsection{Possible Source of High [Fe/H]}

The high [Fe/H] values found for 51 Peg and 55 Cnc require explanation.
  It is too 
much to ascribe to coincidence the presence of two SMR stars in the small
 sample 
of planetary system candidates.  It is also difficult to accept the possibility
 that 55 
Cnc was born with its present photospheric composition given its great age and 
given the scarcity of old metal-rich stars in the Sun's vicinity.  We propose
 instead 
that the original photospheric compositions of 55 Cnc and 51 Peg have been altered 
by the same processes that lead to the creation of the unusual planetary
 systems 
found around these stars.  Our discussion will be limited to a comparison of 51 Peg to the Sun given their close similarities.

Lin et al. (1996) have proposed that 51 Peg B, if it is indeed a gas giant, was 
formed at about 5 AU from the star and at early times (within a few million
 years of 
its formation) migrated inward as a result of interactions with the circumstellar
 disk.  
The disk material (and presumably protoplanets) inside the orbit of 51 Peg B would 
have fallen into the star.  Since much of this material would have been inside
 the so-
called "ice-boundary", much of it would have consisted of refractory elements 
(essentially everything except H and He).  This accreted material would have been 
mixed throughout the convective envelope of the parent star.

Using the present Solar System as a model, we can estimate the effect on the 
photospheric abundances of the Sun had it ingested Mercury, Venus, Earth, and 
Mars in its early history.  Sackmann et al. (1993) calculate that at an age of
 about 30 
Myrs the convective region of the Sun contained about 0.03 M$_{\odot}$.  
Assuming a composition similar to the C1 chondrites (Anders \& Grevesse 1989), 
the addition of the terrestrial planets at this time would have dumped 2.18 x 
$10^{\rm 27}$ g of Fe into the convection zone, leading to an increase of [Fe/H]
 in the photosphere of 0.01 dex.  The addition of 20 M$_{\earth}$, which is
 still only 
0.06 M$_{\rm J}$, would have resulted in an increase of 0.11 dex, a detectable 
change.  The timing of the accretion is critical, as the stellar convection zone
 rapidly 
shrank in size during the first few million years of the Sun's existence; add the 
material too early, and it is diluted throughout a large volume.  The timescale
 for the 
evolution of the protostellar disk is about 5 x $10^{\rm 6}$ yr (Strom et al. 1993).

The next logical question to ask is, Why did the Solar System and the 51 Peg 
systems evolve differently?  After all, both stars are of similar spectral types.
  The 
primary difference seems to be the composition of their photospheres.  If
 51 Peg 
was originally slightly more metal-rich than the Sun, then this would have resulted 
in the formation of a protoplanetary disk more abundant in refractory elements, 
which might have lead to the more rapid formation of protoplanets.  A slightly more 
massive disk may have also lead to greater transfer of mass into the parent
 star 
(Laughlin \& Bodenheimer 1994).  Hence, according to this scenario, a more metal-
rich star is more likely to alter its surface composition during the protostellar
 disk 
evolution phase than a similar but less metal-rich one.  This is consistent with the 
similarity of the 47 UMa system to the Solar System.  The companion of 47 UMa 
orbits at about 2 AU in a low eccentricity orbit (Butler \& Marcy 1996), apparently 
having avoided the orbital decay phenomenon experienced by 51 Peg B.  The 
metallicity of 47 UMa is also less than that of 51 Peg.  Therefore, it appears from 
this small sample that the cutoff value of [Fe/H], below which the drag 
phenomenon will not operate efficiently, is between about +0.1 and +0.2 dex.

This hypothesis can be tested by comparing the compositions of planetary system 
candidates that are in wide-separation binaries.  So far, 55 Cnc is the only such 
system; it is a common proper motion pair (Duquennoy \& Mayor 1991).  
Unfortunately, its companion is an M5 dwarf with m$_{\rm v}$ = 13.2, which 
makes an accurate abundance analysis very difficult.  A careful photometric analysis 
would probably be the best course of action at this time.

If the photospheres of 55 Cnc and 51 Peg are metal-rich relative to their interiors, 
then evolutionary ages derived above are not correct.  The metallicity affects not 
only the effective temperature and radius but also the luminosity.  However, this 
may not fully account for the apparent great age of 55 Cnc.  It was shown in \S 4.2, 
using other age indicators, that 55 Cnc is probably older than 51 Peg or 47 UMa.

\section{Conclusions}

We have undertaken spectroscopic abundance analyses of the five recently 
announced planetary system candidates.  The metallicity is less than solar for HD 
114762, approximately solar for 70 Vir and 47 UMa, and greater than solar
 for 55 
Cnc and 51 Peg.  We have also estimated the projected rotational velocities and 
employed them, along with other data, to estimate the most likely masses of the 
substellar companions of the program stars.  The systems fall into three groups: 51 Peg B (mass $\approx$ 0.51 M$_{\rm J}$) and 55 Cnc B (mass > 0.6 M$_{\rm 
J}$) - both SMR stars with small eccentricities and very small orbital distances, 47 
UMa B at $\approx$ 4.6 M$_{\rm J}$ orbiting at about 2 AU, and 70 Vir B and 
HD 114762 B at > 10 - 20 M$_{\rm J}$ with large eccentricities.

The most surprising, and potentially most important, findings in this study are the 
high metallicities of 55 Cnc and 51 Peg.  These results are consistent with the 
planetary orbital migration model whereby a gas giant migrates to within a few 
hundredths of an AU of the parent star; the material between the planet and the star 
is presumably accreted onto the latter.  If this is confirmed by future observations, 
then Galactic chemical evolution models will have to be revised accordingly.  Also, 
we encourage researchers to direct their planetary search efforts at other SMR 
dwarfs and subgiants.

In order to improve the mass estimates of the companions, future research
 should 
be directed at refining the $v \sin i$ estimates through the use of very high 
resolution spectroscopy.  Improvements in the $M_{\rm V}$ estimates will occur 
when the HIPPARCOS dataset becomes available for the other planet candidates in 
the near future; this will lead to improved estimates of the stellar radii, masses,
 and 
ages.  Additional theoretical work needs to be done on the planet migration process, 
the role of metallicity in the planet formation process, the early evolution of the 
depth of the convection zone in solar-type stars, and stellar models with metal-rich 
envelopes.

\acknowledgments

The author would like to especially thank David L. Lambert and Eric Bakker for 
obtaining high resolution spectra of the program stars at his request and for 
providing helpful suggestions.  Additional helpful comments were given by Jocelyn 
Tomkin and Suchitra Balachandran.  Thanks also go to Geoff Marcy for providing 
a list of his target stars and to Bengt Edvardsson for providing the EW 
measurements of 51 Peg used in the Edvardsson et al. (1993) paper.  This research 
has made use of the Simbad database, operated at CDS, Strasbourg, France.  The 
research has been supported in part by the National Science Foundation (Grant 
AST-9315124) and the Robert A. Welch Foundation of Houston, Texas.

\clearpage

\begin{deluxetable}{lcccr}
\footnotesize
\tablecaption{Atomic Data for Lines Used Here and not
 Listed by Gonzalez \& Lambert (1996).}
\tablewidth{5in}
\tablehead{
\colhead{Spectral line} & \colhead{$\lambda$(\AA)} & \colhead{$\chi_{\rm l}$(eV)} &
\colhead{EW$_\odot$(m\AA)} & \colhead{$\log gf$}
}
\startdata
\ion{C}{1} &6587.61 &8.53 &13.0 &$-$1.20\nl
\ion{O}{1} &7771.94 &9.14 &70.6 &0.25\nl
\ion{O}{1} &7774.16 &9.14 &62.7 &0.13\nl
\ion{O}{1} &7775.38 &9.14 &48.4 &$-$0.11\nl
\ion{Na}{1} &5688.20 &2.10 &121.4 &$-$0.65\nl
\ion{Al}{1} &7835.29 &4.02 &45.0 &$-$0.74\nl
\ion{Al}{1} &7836.11 &4.02 &58.6 &$-$0.57\nl
\ion{Si}{1} &5793.06 &4.93 &44.3 &$-$1.96\nl
\ion{Ca}{1} &5867.55 &2.93 &24.7 &$-$1.62\nl
\ion{Ca}{1} &6455.59 &2.52 &57.3 &$-$1.40\nl
\ion{Cr}{1} &5783.07 &3.32 &32.5 &$-$0.45\nl
\ion{Cr}{1} &5783.87 &3.32 &44.7 &$-$0.21\nl
\ion{Fe}{1} &6089.57 &5.02 &37.0 &$-$0.91\nl
\ion{Fe}{1} &6093.65 &4.61 &31.0 &$-$1.40\nl
\ion{Fe}{1} &7583.78 &3.02 &85.7 &$-$1.95\nl
\ion{Fe}{1} &7586.01 &4.31 &135.3 &$-$0.20\nl
\ion{Fe}{1} &7588.29 &5.03 &28.0 &$-$1.14\nl
\ion{Co}{1} &6454.98 &3.63 &14.5 &$-$0.32\nl
\enddata

\end{deluxetable}

\clearpage

\begin{deluxetable}{lcccrc}
\footnotesize
\tablecaption{Atmospheric Parameters Derived from Fe-Line Analysis.}
\tablewidth{6in}
\tablehead{
\colhead{Star} & \colhead{T$_{\rm eff}$(K)} & \colhead{$\log g$} &
\colhead{$\xi_{\rm t}$(km~s$^{-1}$)} & \colhead{[Fe/H]} & \colhead{N(Fe I,Fe II)}
}
\startdata
55 Cnc &5250 $\pm$ 100 &4.30 $\pm$ 0.05 &1.0 $\pm$ 0.20 &0.24 $\pm$ 0.09 &14, 6\nl
51 Peg &5750 $\pm$ 100 &4.40 $\pm$ 0.10 &1.0 $\pm$ 0.10 &0.22 $\pm$ 0.07 &26, 5\nl
47 UMa &6000 $\pm$ 100 &4.55 $\pm$ 0.05 &1.0 $\pm$ 0.10 &0.10 $\pm$ 0.06 &20, 5\nl
70 Vir &5600 $\pm$ 100 &4.10 $\pm$ 0.10 &1.0 $\pm$ 0.10 &0.04 $\pm$ 0.07 &21, 4\nl
HD 114762 &6000 $\pm$ 100 &4.50 $\pm$ 0.05 &1.0 $\pm$ 0.10 &$-$0.57 $\pm$
 0.06 &17, 3\nl
\enddata

\end{deluxetable}

\clearpage

\begin{deluxetable}{lcccc}
\footnotesize
\tablecaption{Sensitivities of Calculated Abundances to Changes in Model Atmosphere
 Parameters for 51 Peg.}
\tablewidth{0pt}
\tablehead{
\colhead{Line,$\lambda$(m\AA),EW(m\AA)} & \colhead{$\Delta$T$_{\rm eff}$ = $+$250 K} &
\colhead{$\Delta \log g = +0.5$ (cgs)} &
\colhead{$\Delta \xi_{\rm t} = +0.5$ km~s$^{-1}$} &
\colhead{$\Delta$EW$ = +2\%$}
}
\startdata
\ion{O}{1}, 7774.16, 75 &$-$0.24 &$+$0.10 &$-$0.05 &$+$0.10\nl
\ion{Ca}{1}, 6455.59, 70 &$+$0.18 &$-$0.07 &$-$0.10 &$+$0.11\nl
\ion{Sc}{2}, 6604.58, 50 &$-$0.01 &$+$0.21 &$-$0.10 &$+$0.09\nl
\ion{Fe}{1}, 6750.15, 87 &$+$0.21 &$-$0.06 &$-$0.20 &$+$0.16\nl
\ion{Fe}{1}, 7588.29, 39 &$+$0.10 &$-$0.02 &$-$0.05 &$+$0.07\nl
\ion{Fe}{2}, 6432.68, 47 &$-$0.11 &$+$0.22 &$-$0.10 &$+$0.11\nl
\enddata

\end{deluxetable}

\begin{table}
\dummytable
\end{table}

\clearpage

\begin{deluxetable}{lcccccccc}
\footnotesize
\tablecaption{Values of T$_{\rm eff}$ for Program Stars from other Sources.}
\tablewidth{0pt}
\tablehead{
\colhead{Star} & \colhead{$\beta$} & \colhead{$b-y$} &
 \colhead{T$_{\rm eff}$(Ed. et al.)\tablenotemark{a}} &
\colhead{T$_{\rm eff}(b-y)$\tablenotemark{a}} &
\colhead{T$_{\rm eff}(\beta)$\tablenotemark{a}} &
\colhead{$<$T$_{\rm eff}>$\tablenotemark{b}} &
\colhead{$\log g$\tablenotemark{c}} &
\colhead{[Fe/H]\tablenotemark{c}}
}
\startdata
51 Peg & 2.603 & 0.416 & 5755 & 5725 & 5889 & 5779 & 4.44 & 0.24\nl
47 UMa & 2.606 & 0.392 & 5882 & 5850 & 5923 & 5943 & 4.45 & 0.09\nl
70 Vir & 2.576 & 0.446 & \nodata & 5522 & 5578 & 5575 & 4.05 & 0.03\nl
HD 114762 & 2.590 & 0.363 & 5871 & 5887 & 5742 & 5907 & 4.39 & $-$0.61\nl
\enddata
\tablenotetext{a}{T$_{\rm eff}(b-y)$ and T$_{\rm eff}$ were estimated from the
Str\"omgren photometry given by Hauck \& Mermilliod (1980) and using Equation 
1.  Ed. et al. refers to Edvardsson et al. (1993).}
\tablenotetext{b}{The mean value of T$_{\rm eff}$ was calculated from the two 
photometric estimates, with half weight given to each, and the spectroscopic 
estimates of this study.}
\tablenotetext{c}{These new $\log g$ and [Fe/H] estimates are based on 
$<$T$_{\rm eff}>$.}

\end{deluxetable}

\clearpage

\begin{deluxetable}{lccccccccccc}
\scriptsize
\tablecaption{Final Adopted Abundances for the Program Stars.}
\tablewidth{0pt}
\tablehead{
\colhead{Element} & \colhead{$\log \epsilon_{\odot}$} &
 \colhead{55 Cnc} & \colhead{} &
 \colhead{51 Peg} & \colhead{} & \colhead{47 UMa} &
\colhead{} & \colhead{70 Vir} & \colhead{} & \colhead{HD 114762} &
\colhead{}\\
\colhead{} & \colhead{} & \colhead{[X/H]} & \colhead{N} &
\colhead{[X/H]} & \colhead{N} &
\colhead{[X/H]} & \colhead{N} &
\colhead{[X/H]} & \colhead{N} &
\colhead{[X/H]} & \colhead{N}
}
\startdata
Li & 1.06 & $<$0.56 & 1 & 1.50$\pm$0.11 & 1 & 1.90$\pm$0.07 &
 1 & 1.85$\pm$0.07 & 1 & 2.77$\pm$0.07 & 1\nl
C & 8.55 & 0.22$\pm$0.15 & 1 & 0.12$\pm$0.12 & 2 & $-$0.05$\pm$0.12 &
 2 & $-$0.05$\pm$0.12 & 2 & $-$0.54$\pm$0.11 & 1\nl
O & 8.94 & \nodata & \nodata & 0.24$\pm$0.11 & 3 & 0.10$\pm$0.11 &
 3 & $-$0.07$\pm$0.11 & 3 & \nodata & \nodata\nl
Na & 6.34 & 0.33$\pm$0.13 & 2 & 0.18$\pm$0.09 & 3 & 0.20$\pm$0.09 &
 1 & 0.04$\pm$0.10 & 1 & $-$0.40$\pm$0.09 & 1\nl
Mg & 7.61 & \nodata & \nodata & 0.15$\pm$0.11 & 1 & 0.04$\pm$0.09 &
 1 & $-$0.01$\pm$0.11 & 1 & $-$0.38$\pm$0.09 & 1\nl
Al & 6.51 & 0.52$\pm$0.12 & 1 & 0.29$\pm$0.08 & 3 & 0.01$\pm$0.07 &
 2 & 0.11$\pm$0.09 & 2 & \nodata & \nodata\nl
Si & 7.58 & 0.39$\pm$0.14 & 1 & 0.21$\pm$0.11 & 3 & 0.08$\pm$0.07 &
 2 & $-$0.01$\pm$0.08 & 2 & $-$0.41$\pm$0.07 & 2\nl
S & 7.29 & \nodata & \nodata & 0.22$\pm$0.13 & 1 & 0.09$\pm$0.11 &
 1 & \nodata & \nodata & \nodata & \nodata\nl
Ca & 6.37 & 0.30$\pm$0.10 & 2 & 0.16$\pm$0.09 & 3 & 0.13$\pm$0.08 &
 2 & 0.03$\pm$0.08 & 3 & $-$0.44$\pm$0.09 & 1\nl
Sc & 3.11 & \nodata & \nodata & 0.33$\pm$0.07 & 2 & 0.18$\pm$0.05 &
 2 & 0.13$\pm$0.07 & 2 & $-$0.45$\pm$0.05 & 2\nl
Ti & 4.96 & 0.25$\pm$0.09 & 2 & 0.24$\pm$0.07 & 4 & 0.21$\pm$0.06 &
 2 & 0.12$\pm$0.08 & 4 & $-$0.28$\pm$0.05 & 4\nl
Cr & 5.71 & 0.34$\pm$0.09 & 2 & 0.25$\pm$0.08 & 2 & 0.12$\pm$0.07 &
 2 & 0.07$\pm$0.08 & 2 & $-$0.70$\pm$0.08 & 1\nl
Fe & 7.53 & 0.24$\pm$0.09 & 20 & 0.22$\pm$0.07 & 31 & 0.10$\pm$0.06 &
 25 & 0.04$\pm$0.07 & 25 & $-$0.57$\pm$0.06 & 20\nl
Co & 4.93 & 0.49$\pm$0.11 & 1 & 0.30$\pm$0.08 & 2 & \nodata &
 \nodata & 0.00$\pm$0.09 & 1 & \nodata & \nodata\nl
Ni & 6.27 & 0.29$\pm$0.16 & 1 & 0.31$\pm$0.08 & 2 & 0.15$\pm$0.07 &
 1 & 0.03$\pm$0.09 & 1 & $-$0.51$\pm$0.06 & 2\nl
Y & 2.25 & \nodata & \nodata & \nodata & \nodata & 0.04$\pm$0.07 &
 1 & $-$0.12$\pm$0.09 & 1 & $-$0.66$\pm$0.06 & 2\nl
Ba & 2.23 & 0.14$\pm$0.12 & 1 & 0.17$\pm$0.10 & 1 & 0.20$\pm$0.07 &
 1 & $-$0.02$\pm$0.09 & 1 & $-$0.76$\pm$0.07 & 1\nl
\enddata
\tablecomments{The uncertainties in [X/H] are a combination of the random
and systematic (uncertainties in atmospheric parameters) error sources.  The 
lithium abundances are given as $\log \epsilon$, rather than [X/H].}

\end{deluxetable}

\clearpage

\begin{deluxetable}{lcc}
\footnotesize
\tablecaption{Measured Values of Macroturbulent and Rotational Line 
Broadening.}
\tablewidth{5in}
\tablehead{
\colhead{Star} & \colhead{$\zeta_{\rm RT}$} &
\colhead{$v \sin i$}\\
\colhead{~~Fe I line} & \colhead{km~s$^{-1}$} &
\colhead{km~s$^{-1}$}
}
\startdata
Sun & &\nl
~~5379.586 & 3.55$\pm$0.05 & $\equiv$1.9\nl
~~5638.249 & 3.55$\pm$0.05 & $\equiv$1.9\nl
~~5379.586 & 3.60$\pm$0.10 & 1.9$\pm$0.1\nl
~~5638.249 & 3.50$\pm$0.10 & 2.0$\pm$0.1\nl
55 Cnc & &\nl
~~5379.586 & 2.7$\pm$0.3 & 1.2$\pm$0.7\nl
~~5638.249 & 2.2$\pm$0.4 & 1.9$\pm$0.7\nl
51 Peg & &\nl
~~6213.435 & 3.50$\pm$0.30 & 2.0$\pm$0.5\nl
~~6322.691 & 4.10$\pm$0.30 & 1.5$\pm$0.7\nl
47 UMa & &\nl
~~5379.586 & 4.10$\pm$0.20 & 1.9$\pm$0.6\nl
~~5638.249 & 3.95$\pm$0.20 & 1.9$\pm$0.7\nl
70 Vir & &\nl
~~5379.586 & 4.05$\pm$0.05 & 0.4$\pm$0.4\nl
~~5638.249 & 4.10$\pm$0.10 & 0.5$\pm$0.5\nl
HD 114762 & &\nl
~~5379.586 & 4.1$\pm$0.1 & 1.0$\pm$1.0\nl
~~5638.249 & 4.1$\pm$0.1 & 1.0$\pm$0.6\nl
\enddata

\end{deluxetable}

\clearpage

\begin{deluxetable}{lcccccc}
\footnotesize
\tablecaption{Age and Mass Estimates of Program Stars Using VandenBerg
Evolutionary Tracks.}
\tablewidth{6in}
\tablehead{
\colhead{Star} & \colhead{$M_{\rm V}$} & \colhead{$M_{\rm V}$} &
 \colhead{$M_{\rm V}$} &
\colhead{Age (Gyr)} & \colhead{Mass (M$_{\odot}$)} & \colhead{Age (Gyr)}\\
\colhead{} & \colhead{(phot.)} & \colhead{(astrom.)} & \colhead{(adopted)} &
\colhead{(M$_{\rm v}-$T$_{\rm eff}$)} &
\colhead{} & \colhead{(Ed. et al.)}
}
\startdata
55 Cnc & \nodata & $5.28^{+0.08}_{-0.07}$ & 5.28$\pm$0.08 &
13$\pm$2 & 0.90$\pm$0.02 & \nodata\nl
51 Peg & 3.7$\pm$0.2 & 4.56$\pm$0.04 & 4.56$\pm$0.04 &
$3^{+2}_{-1}$ & 1.13$\pm$0.03 & 8.5\nl
47 UMa & 4.1$\pm$0.2 & 4.31$\pm$0.03 & 4.31$\pm$0.03 &
5$\pm$1 & 1.10$\pm$0.02 & 6.9\nl
70 Vir & 3.7$\pm$0.2 & 3.71$\pm$0.04 & 3.71$\pm$0.04 &
7.0$\pm$0.5 & 1.20$\pm$0.01 & \nodata\nl
HD 114762 & 4.6$\pm$0.2 & \nodata & 4.6$\pm$0.20 &
17$\pm$3 & 0.79$\pm$0.02 & 13.8\nl
\enddata

\end{deluxetable}

\clearpage

\figcaption{Portions of the spectra of 51 Peg and the Sun (reflected off
 the asteroid Vesta) containing the Li I line and two Fe I lines of moderate strength.
  The spectra were obtained with the same instrument (on different dates); the resolutions have been matched by smoothing the (slightly) higher resolution spectrum.  The Sun and 
51 Peg have nearly identical physical parameters; hence, the stronger lines in the 
spectrum of 51 Peg are due to a greater Fe abundance.}

\figcaption{The Fe I (filled circles) and Fe II (plus signs) abundances calculated
 using the model parameters given in Table 2 are plotted versus the
 lower excitation potential ($\chi_{\rm l}$) and $\log \left(EW/\lambda\right)$.  The solid lines are least-squares fits through the Fe I data points.  The dotted lines in the first diagram are the least-squares fits when $T_{\rm eff}$ is changed by $\pm$ 250 K.  The 
dotted lines in the second diagram are the least-squares fits when $\xi_{\rm t}$ is 
changed by $\pm$ 0.2 km~s$^{-1}$.  Even though the lower dotted line in the 
second diagram (corresponding to $\xi_{\rm t}$ = 1.2 km~s$^{-1}$) appears to be 
a better solution, it was not chosen because the slopes of the least-squares fits have 
been affected by the two strong Fe I lines near $\log \left(EW/\lambda\right)$ = 
-4.7.  These two points do not affect the $T_{\rm eff}$ estimate significantly since 
they have very different values of $\chi_{\rm l}$.}

\figcaption{Best-fit synthetic profile (solid curve) of the Fe I
 line at 5638.28 \AA\ in the spectrum of 70 Vir using $\zeta_{\rm RT}$ = 4.1
 and $v \sin i$ = 0.5 km~s$^{-1}$; the dotted curve corresponds to
 $\zeta_{\rm RT}$ = 4.2 and $v \sin i$ = 0.0 km~s$^{-1}$ and the dashed curve
 to $\zeta_{\rm RT}$ = 3.8 and $v \sin i$ = 1.4 km~s$^{-1}$.  Also shown are the
 residual differences between the observations and the synthesis; the large low frequency fluctuations in this plot are due to a slight asymmetry in the observed line profile.  Although all three residual plots appear very similar to visual inspection, the best
 fit was chosen in a quatitative manner as the one with the smallest standard deviation.}

\figcaption{Histogram of the distribution of photometric [Fe/H] estimates
 for F5 to G2 dwarfs within 80 pc of the Sun determined by Marsakov \& Shevelev (1995).  
The average [Fe/H] value is indicated by a solid line and the program stars by 
dotted lines.  The [Fe/H] values of the program stars are from this study.}

\clearpage

\end{document}